\documentclass{article}
\usepackage{booktabs}
\usepackage{multirow}
\usepackage{amsmath}
\usepackage{spconf,amsmath,graphicx,hyperref}
\usepackage{amssymb}

\usepackage{cite}

\title{Poly-SVC: Polyphony-Aware Singing Voice Conversion with Harmonic Modeling}
\name{
  Chen Geng\textsuperscript{1,3,*}, 
  Meng Chen\textsuperscript{2,*}, 
  Ruohua Zhou\textsuperscript{1,3,$\dagger$}, 
  Ruolan Liu, 
  Weifeng Zhao\textsuperscript{2}
  \thanks{\parbox[t]{8cm}{
  $\dagger$ Corresponding author.\\
  * Equal contribution.
    }
  }
}

\address{\textsuperscript{1}School of Intelligence Science and Technology,\\ Beijing University of Civil Engineering and Architecture, Beijing, China\\ 
\textsuperscript{2}Lyra Lab, Tencent Music Entertainment, Shenzhen, China\\
\textsuperscript{3}Beijing Key Laboratory of Super Intelligent Technology for Urban Architecture, Beijing, China}
\begin{document}
\ninept
\maketitle
\begin{abstract}
Singing Voice Conversion (SVC) aims to transform a source singing voice into a target singer while preserving lyrics and melody. Most existing SVC methods depend on F0 extractors to capture the lead melody from clean vocals. However, no existing method can reliably extract clean vocals from accompanied recordings without leaving residual harmonies behind. In this paper, we innovatively propose Poly-SVC, a zero-shot, cross-lingual singing voice conversion system designed to process residual harmonies. Poly-SVC is composed of three key components: a Constant-Q Transform (CQT)-based pitch extractor to preserve both the lead melody and residual harmony, a random sampler to reduce interference information from the CQT and a diffusion decoder based on Conditional Flow Matching (CFM) that fuses pitch, content, and timbre features into natural-sounding polyphonic outputs. Experiments demonstrate that Poly-SVC surpasses the baseline models in naturalness, timbre similarity and harmony reconstruction across both harmony-rich and single-melody recordings.
\end{abstract}
\begin{keywords}
Singing voice conversion, polyphonic vocal, residual harmony
\end{keywords}
\section{Introduction}
\label{sec:intro}

Singing voice conversion (SVC) is an emerging research hotspot that converts one singer’s vocal identity and style to sound like another while keeping the original lyrics, melody, and various vocal techniques\cite{ferreira2025freesvc,bai2024spa,liu2024zero,zhou2025syki}. The task addressed in this work presents greater challenges than conventional SVC, as it deals with the mismatch between clean training data and real-world inputs, where vocals are often mixed with accompaniment and cannot be perfectly isolated. 

Most state-of-the-art SVC systems adopt a feature disentanglement paradigm, which decomposes the singing voice into three primary components: timbre, content, and pitch \cite{ferreira2025freesvc,bai2024spa,liu2024zero,zhou2025syki}. To support this framework, recent studies have leveraged powerful pre-trained models for each aspect. For content modeling, self-supervised learning (SSL) models such as wav2vec \cite{baevski2020wav2vec} and HuBERT \cite{hsu2021hubert}, as well as automatic speech recognition (ASR) models such as Whisper \cite{radford2023robust}, are widely adopted to provide robust, high-resolution linguistic representations, greatly improving intelligibility in the synthesized output. At the same time, F0-based pitch estimation methods,  including RMVPE \cite{wei2023rmvpe} and Crepe \cite{kim2018crepe}, are commonly used for capturing the melodic contour. For timbre representation, both speaker embeddings and prompt-based timbre encodings are widely adopted. On one hand, speaker verification models such as CampPlus \cite{wang2023cam++} extract stable and discriminative voice embeddings; on the other hand, approaches like F5-TTS \cite{chen2024f5} utilize masking-based encoders to derive timbre features directly from audio prompts. Both paradigms have proven effective in preserving voice identity. In addition, the Diffusion Transformer (DiT) and VITS architectures have demonstrated excellent performance in high-fidelity speech synthesis and singing voice conversion \cite{chen2024f5,wang2024maskgct,liu2021fastsvc}. 

Although current methods deliver strong results in ideal conditions with clean vocals, they fail to bridge the gap between such settings and real-world data. In real-world applications, clean vocals are often mixed with background accompaniment and embedded harmonies, which makes it challenging to isolate a single dominant pitch. This mismatch severely limits the effectiveness of F0-based pitch extractors in real-world singing scenarios, often resulting in off-key artifacts and degraded audio quality.

Currently, there are two main approaches to pitch extraction in real-world conditions. The first approach performs vocal separation using tools like Demucs \cite{rouard2023hybrid} to get the clean singing voice before applying F0 estimation. However, in most cases, vocal separation is imperfect and leaves behind residual harmonies, which severely interfere with F0 extractors and degrade pitch accuracy. Even with clearly isolated vocals, certain singing techniques still lead to distorted F0 predictions. The second approach directly applies F0 extraction to the raw audio with accompaniment \cite{wei2023rmvpe} and \cite{ chen2025singing}, which often leads to pitch tracking errors due to interference from instruments and overlapping harmonics. Hence, the pitch extractor is the critical determinant of SVC performance in real-world scenarios. 

To address the gap between ideal conditions and real-world scenarios, we propose Poly-SVC, a singing voice conversion system specifically designed to tackle the pitch prediction challenges caused by residual harmonies introduced during vocal-accompaniment separation, as discussed in the first approach. Our contributions include:
(1) Introducing a polyphonic-aware pitch extractor that leverages Constant-Q Transform (CQT) features to preserve both the lead melody and residual harmony essential for robust pitch modeling; 
(2) Designing a random sampler to enhance the CQT-based pitch extractor, leveraging limited MIDI-labeled data, effectively suppressing non-pitch components in CQT features and reducing information leakage in pitch modeling; 
(3) Building an end-to-end SVC framework with a Conditional Flow Matching (CFM)-based diffusion decoder \cite{FLOWMATCHINGLipmanCBNL23} that integrates advanced content, pitch, and timbre features for high-fidelity, harmony-preserving zero-shot singing voice conversion.
The rest of this paper is organized as follows: Section II presents the proposed Poly-SVC system, experiments are described in Section III and Section IV concludes this paper.

\begin{figure*}[htb]
\centering
\begin{minipage}[b]{.9\linewidth}
  \centerline{\includegraphics[width=\linewidth]{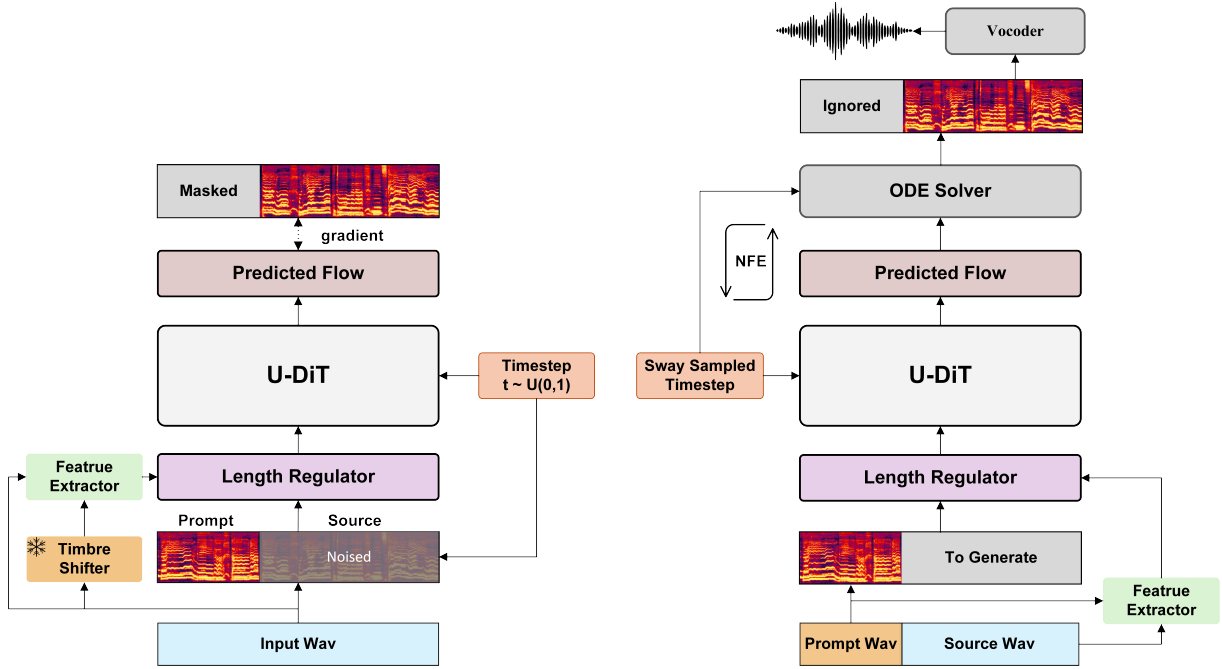}}
\end{minipage}
\caption{ Comprehensive diagram of the Poly-SVC model. The left shows the training procedure, the right the inference procedure. Snowflake represents the
parameter that remains unchanged when training the SVC framework.}
\label{fig:framework}
\end{figure*}

\section{Methods}
\label{sec:format}
Fig.~\ref{fig:framework} illustrates an overview of our Poly-SVC framework. Following prior SVC methods \cite{chen2024f5} and \cite{liu2021fastsvc}, we first extract the mel-spectrogram as the acoustic representation and apply a Timbre Shifter based on OpenVoice \cite{qin2023openvoice} to align the distributions between training and inference, thereby reducing the timbre leak from the content representation. A feature extractor with Random Sampler then encodes content, pitch, and timbre features, which are aligned with mel-spectrogram by a learnable length regulator. These fused representations are then used to condition a CFM-based diffusion decoder trained with a CFM loss, with the aim of reconstructing clean mel-spectrograms. Finally, we finetune a pretrained Firefly-GAN \cite{liao2024fish} vocoder on our harmony dataset to convert the output spectrograms into high-fidelity audio. In the following sections, we provide detailed descriptions of the Feature Extractor, the Random Sampler, and the CFM-based Singing Voice Converter.

\begin{figure*}[htb]
\centering
\begin{minipage}[b]{.91\linewidth}
  \centerline{\includegraphics[width=\linewidth]{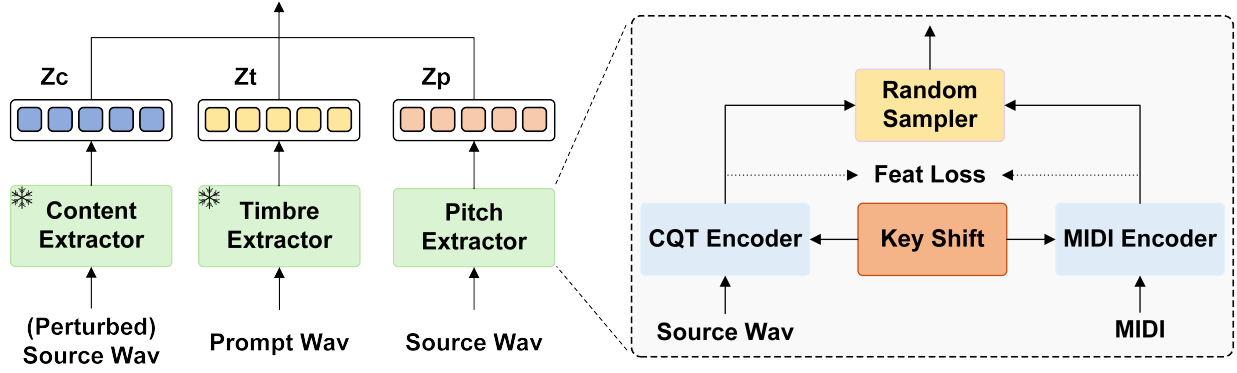}}
\end{minipage}
\caption{Feature extractor framework. Snowflake represents the parameter that remains unchanged when training the SVC framework.}
\label{fig:random_sampler}
\end{figure*}

\subsection{Feature Extractor}
\label{ssec:subhead}
As illustrated in Fig.~\ref{fig:random_sampler}, the Feature Extractor integrates three distinct features: content $z_c$, pitch $z_p$, and timbre $z_t$. Content and timbre features are obtained via pretrained, open-source models, such as Whisper \cite{radford2023robust} and CampPlus \cite{wang2023cam++}. In contrast, accurate pitch extraction presents additional challenges in real-world scenarios. To address the challenge of extracting accurate pitch information from vocals with overlapping residual harmonies, we adopt the CQT spectrogram as the pitch feature representation, due to its capacity to preserve both the lead melody and accompanying harmonies. The CQT spectrogram maintains a constant ratio between center frequency and bandwidth, resulting in uniform bin distribution per octave and offering high resolution for pitch structures across multiple octaves. Due to its pitch-aligned structure and logarithmic frequency scaling, this representation provides a strong foundation for music-related analysis and has been widely applied in tasks such as music information retrieval (MIR). Its ability to simultaneously capture spectral and temporal structures across octaves makes it well-suited for harmony modeling, particularly in polyphonic vocal scenarios where multiple pitches co-occur and change dynamically over time. In contrast to traditional F0-based methods, which often fail in polyphonic scenarios, our pitch extractor leverages the CQT-based representation for more reliable and comprehensive pitch modeling. 
\subsection{Random Sampler}
\label{ssec:subhead}
The CQT-based representation encodes the full melodic content in a logarithmic frequency scale, but it also introduces extraneous information such as timbre. To mitigate the impact of irrelevant information, we crop the CQT spectrogram to the typical human singing range of 32Hz-1kHz, discarding out-of-range components. We then utilize a small set of MIDI-aligned data to supervise training. As shown in Fig.~\ref{fig:random_sampler}, the audio input is transformed into a CQT matrix and encoded into  $z_{cqt}$  via a CQT Encoder. When MIDI annotations are available, parallel MIDI-labelled data are encoded by a MIDI encoder. The CQT-based representation encodes the full melodic content in a logarithmic frequency scale, but it also introduces extraneous information such as timbre. To mitigate the impact of irrelevant information, we crop the CQT spectrogram to the typical human singing range of 32Hz-1kHz, discarding out-of-range components. We then utilize a small set of MIDI-aligned data to supervise training. As shown in Fig.~\ref{fig:random_sampler}, the audio input is transformed into a CQT matrix and encoded into  $z_{cqt}$ via a CQT Encoder. When MIDI annotations are available, parallel MIDI-labelled data are encoded by a MIDI encoder.
\begin{equation}
\mathcal{L}_{RS} = \left\| E_{CQT}(CQT(x)) - E_{MIDI}(MIDI) \right\|_1
\end{equation}

During inference, the CQT-based path enables robust extraction of multiple simultaneous melody lines without relying on MIDI inputs, naturally supporting harmony preservation in complex vocal scenarios. Moreover, simple shifting of the CQT matrix along the frequency axis allows pitch transposition without additional processing.
\subsection{CFM-based Singing Voice Convertor}
\label{ssec:subhead}

Within the CFM-based Singing Voice Convertor, the model first takes a pair of samples $\left(x_0,x_1\right)$, where $x_0$ is sampled from a Gaussian distribution $p_0$, and $x_1$ is the target mel-spectrogram. The interpolation $\psi_t\left(x_0,x_1\right)=\left(1-t\right)x_0+tx_1$ defines a linear flow path from source to target. The model is trained to predict the velocity field $v_t\left(\psi_t\left(x_0,x_1\right),c\right)$, where $c$ denotes the conditioning information, including the extracted features $z$ and mel-spectrograms. Here, $\theta$ represents the learnable parameters of the U-DiT module that control the velocity estimation process. The training objective minimizes the squared error between the predicted velocity and the ground truth velocity of the path.

\begin{table*}[t]
\centering
\caption{Comparison of MOS and SIM-MOS on Single-Melody and Harmony.}
\label{tab:mos_results}
\begin{tabular}{lccccc}
\toprule
\multirow{2}{*}{\textbf{Approach}} & 
\multicolumn{2}{c}{\textbf{Single-Melody}} & 
\multicolumn{2}{c}{\textbf{Harmony}} \\
\cmidrule(lr){2-3} \cmidrule(lr){4-5}
 & \textbf{MOS} & \textbf{SIM-MOS} & \textbf{MOS} & \textbf{SIM-MOS} \\
\midrule
Ground Truth            & 4.12 \text{\( \pm \)} 0.11 & -                & 3.92 \text{\( \pm \)} 0.11 & - \\
so-vits-svc             & 3.57 \text{\( \pm \)} 0.14 & 3.15 \text{\( \pm \)} 0.13  & 1.64 \text{\( \pm \)} 0.10 & 2.08 \text{\( \pm \)} 0.09 \\
DDSP-SVC                & 3.83 \text{\( \pm \)} 0.13 & 3.33 \text{\( \pm \)} 0.11  & 2.98 \text{\( \pm \)} 0.11 & 2.82 \text{\( \pm \)} 0.10 \\
SeedVC                  & 3.85 \text{\( \pm \)} 0.13 & 3.74 \text{\( \pm \)} 0.10  & 3.35 \text{\( \pm \)} 0.12 & 3.40 \text{\( \pm \)} 0.08 \\
\midrule
Poly-SVC (w/o TS)       & 3.96 \text{\( \pm \)} 0.13 & 3.66 \text{\( \pm \)} 0.11  & 3.71 \text{\( \pm \)} 0.10 & 3.32 \text{\( \pm \)} 0.08 \\
Poly-SVC (w/o RS)       & 3.92 \text{\( \pm \)} 0.13 & 3.71 \text{\( \pm \)} 0.12  & 3.62 \text{\( \pm \)} 0.13 & 3.36 \text{\( \pm \)} 0.09 \\
\textbf{Poly-SVC}       & \textbf{3.98 \text{\( \pm \)} 0.12} & \textbf{3.78 \text{\( \pm \)} 0.11} & \textbf{3.75 \text{\( \pm \)} 0.10} & \textbf{3.42 \text{\( \pm \)} 0.09} \\
\bottomrule
\end{tabular}
\end{table*}

\begin{equation}
\begin{split}
\mathcal{L}_{\text{CFM}}(\theta) &= \mathbb{E}_{t, q(x_1), p(x_0)} \\
&\quad \left\| v_t(\psi_t(x_0, x_1), c) - \frac{d}{dt} \psi_t(x_0, x_1) \right\|^2 
\end{split}
\end{equation}

In inference, as shown in Fig.~\ref{fig:framework}, to transform the singing voice of a source singer into the timbre of a prompt singer while preserving the other features, we first employ a feature extractor to extract features $z_{src}$ and $z_{ref}$ from the source and prompt singing voice, respectively. At the same time, we obtain the mel-spectrogram $x_{ref}$ from the prompt. All representations are temporally aligned via a length regulator and subsequently fused to guide singing voice conversion during synthesis. The DiT module takes the fused features as input and predicts the continuous-path vector $v_t$. Then, an ODE solver with sway sample timestep infers the predicted mel-spectrogram, which is finally converted back into waveform audio via a vocoder.

\begin{equation}
v_t(\psi_t(x_0), c) = v_t\left((1-t)x_0 + tx_1 \mid x_{ref}, z_{ref}, z_{src}\right)
\end{equation}

The number of function evaluations (NFE) corresponds to how many times the network is invoked when using multiple flow-step values between 0 and 1 to approximate the integration.

\begin{equation}
f_{\text{sway}}(u; s) = u + s \cdot \left( \cos\left(\frac{\pi}{2} u\right) - 1 + u \right)
\end{equation}

\section{Experimemts}
\label{sec:pagestyle}

\subsection{Dataset}
\label{ssec:subhead}
We use a wide variety of datasets covering both speech and singing, encompassing multiple languages, audio durations, and speaker counts. For speech data, we adopt the Emilia dataset \cite{he2024emilia}, a 101k-hour multilingual speech corpus rich in expressive speaking styles, which provides a robust foundation for modeling natural speech. A small subset is sampled for regular voice conversion training. For singing data, we utilize m4singer \cite{zhang2022m4singer}, OpenSinger \cite{huang2021opensinger}, OpenCpop \cite{wang2022opencpop}, PopBuTFy \cite{PopBuTFyLiu00ZZ22} and VocalSet \cite{wilkins2018vocalset}, which contain English and Chinese singing data with clean, single-melody vocals. The m4singer \cite{zhang2022m4singer} additionally includes a subset with MIDI annotations. 

Notably, as no suitable open-source dataset provides ground-truth vocals paired with harmony, we simulate real-world scenarios by extracting vocal tracks directly from accompanied full-mix songs using the Ultimate Vocal Remover (UVR)\footnote{https://github.com/Anjok07/ultimatevocalremovergui}. To support training and evaluation under harmony-rich conditions, we process 70 hours of publicly available singing recordings in English, Chinese, Cantonese, and Japanese. The extracted vocals are further dereverberated to enable clean vocal supervision in polyphonic settings.

For evaluation, we selected 20 source audio samples from the aforementioned datasets, including 10 single-melody samples covering both singing and speech, and 10 polyphonic singing samples spanning Chinese, English, Cantonese, and Japanese, with each audio clip ranging in duration from 5 to 15 seconds. Additionally, one male and one female speaker were selected from the PopBuTFy \cite{PopBuTFyLiu00ZZ22} dataset as target timbres for inference. The remaining recordings were used for training.

\subsection{Experimental setups}
\label{ssec:subhead}
We adopt whisper-small\footnote{https://huggingface.co/openai/whisper-small} for content extraction. CampPlus\footnote{ https://huggingface.co/funasr/campplus} for timbre extraction, and the U-DiT from SeedVC \cite{liu2024zero} for DiT module. The Timbre Shifter is based on OpenVoice \cite{qin2023openvoice}, while the vocoder relies on Firefly-GAN \cite{liao2024fish}, which we subsequently fine-tune on our harmony dataset. For pitch extraction, the audio is first resampled to 44.1kHz. A CQT matrix is then computed using a hop length of 441, with 12 bins per octave and a total of 84 bins. Both the CQT encoder and the MIDI encoder are implemented as multi-layer transformers. The optimizer is AdamW with peak learning rate of 1e-4, exponentially decays to a minimum of 1e-5.

To evaluate the effectiveness of our model, we compare Poly-SVC with open-sourced baselines: so-vits-svc\footnote{https://github.com/svc-develop-team/so-vits-svc}, DDSP-SVC5\footnote{ https://github.com/yxlllc/DDSP-SVC
}, and the SeedVC \cite{liu2024zero} system. As no extant objective metric adequately captures the quality of conversion, especially in harmony-rich scenarios, subjective evaluation is the primary method of assessment. To ensure a reliable subjective evaluation, we recruited 12 Chinese participants fluent in English and with basic knowledge of music theory. Each participant evaluated the audio using the standard 5-level Mean Opinion Score (MOS) and Similarity-MOS (SIM-MOS) to quantify perceived timbre and melodic fidelity across baselines. We employed a calibrated scale with graded criteria for each level, defining 3 as the perceptibility threshold; scores below this level indicate pronounced acoustic distortion in MOS and clear timbre deviation in SIM-MOS. The process lasted about two hours with scheduled rest intervals, and all individual ratings were retained for analysis. This dual-MOS framework provides a more reliable assessment.

\subsection{Experimental Results}
\label{ssec:subhead}
As shown in Table~\ref{tab:mos_results}, our model achieves state-of-the-art performance in the polyphonic condition, substantially outperforming all baselines by effectively maintaining the harmonic structure and richness of the original input. In the single-melody scenario, it also surpasses existing baselines, primarily due to significant improvements in handling special expressions such as glottal fry.

\begin{figure}[htb]

\begin{minipage}[b]{.32\linewidth}
  \centering
  \centerline{\includegraphics[width=\linewidth]{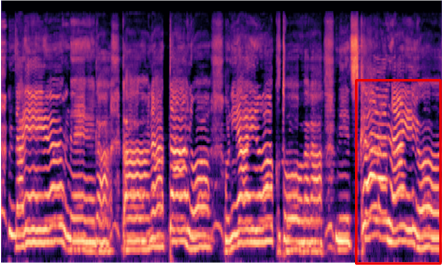}}
  \centerline{(a) Ground Truth}\medskip
\end{minipage}
\hfill
\begin{minipage}[b]{.32\linewidth}
  \centering
  \centerline{\includegraphics[width=\linewidth]{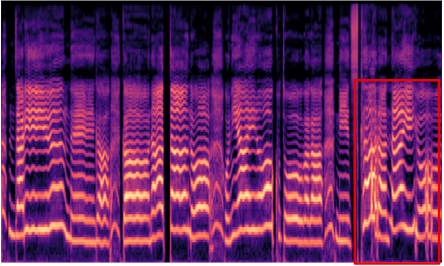}}
  \centerline{(b) SeedVC}\medskip
\end{minipage}
\hfill
\begin{minipage}[b]{0.32\linewidth}
  \centering
  \centerline{\includegraphics[width=\linewidth]{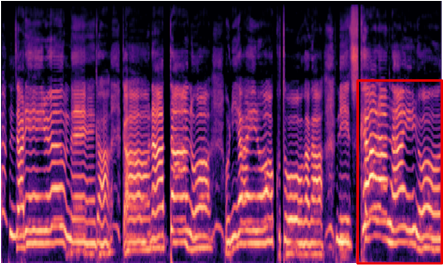}}
  \centerline{(c) Poly-SVC (Ours)}\medskip
\end{minipage}
\caption{Spectrogram comparison of the baselines model under polyphonic conditions.}
\label{fig:mel_compare}
\end{figure}

To further demonstrate the effectiveness of our model under polyphonic conditions, we present a qualitative comparison in Fig.~\ref{fig:mel_compare}. For comparison, we select SeedVC \cite{liu2024zero} as it outperforms other baselines under polyphonic conditions. As illustrated in Fig.~\ref{fig:mel_compare} (a), the input mel-spectrogram comprises multiple overlapping melodic lines. SeedVC \cite{liu2024zero}, captures only the primary vocal melody while failing to preserve the underlying harmonic structure. In contrast, Poly-SVC successfully preserves both the main melody and harmonic components.  In addition to its inability to preserve harmonic components, SeedVC \cite{liu2024zero} frequently makes errors even in the extraction of the primary melody. As illustrated in the red-boxed region, SeedVC \cite{liu2024zero} exhibits a significant pitch prediction error, resulting in noticeable spectral distortion. In contrast, our model not only reconstructs the lead melody but also retains the underlying harmonic structure with greater fidelity.

To validate the contribution of each component further, we conducted ablation experiments by removing the Random Sampler (RS) and the Timbre Shifter (TS) in turn.  Although TS has demonstrated strong performance under clean, single-melody conditions, it remains essential to evaluate its effectiveness in more challenging polyphonic scenarios. Removing TS leads to a marked drop in SIM-MOS, indicating that TS effectively suppresses timbre leakage from content representations and ensures better alignment between training and inference conditions. In contrast, removing RS results in a consistent decline in both MOS and SIM-MOS scores, suggesting that RS plays a key role in guiding the pitch extractor to focus on melody-relevant cues in CQT features, thereby reducing pitch-irrelevant noise and preventing timbre artifacts. Overall, integrating both the Random Sampler and Timbre Shifter yields the highest perceptual quality, with notable improvements in timbre consistency.

\section{Conclusion}
\label{sec:typestyle}
This study highlights the significant challenges inherent in real-world singing voice conversion, particularly due to the challenge of obtaining clean singing vocals. To address these issues, we proposed Poly-SVC, a singing voice conversion framework designed for real-world scenarios where vocal-accompaniment separation often leaves residual harmonies. Poly-SVC employs a CQT-based pitch extractor for full melodic modeling and a Random Sampler to mitigate information leakage during training, while integrating content, timbre, and pitch representations into a CFM-based diffusion decoder for high-fidelity voice conversion with improved naturalness, timbre similarity, and harmonic reconstruction.

Future work will focus on addressing content overlapping in
singing voice conversion, as current methods fail to adequately represent overlapping content.


\bibliographystyle{IEEEtran}
\bibliography{strings,refs}

\end{document}